\documentclass[11pt, a4paper]{article}%Delete onecolumn and 11pt later.

\usepackage{jheppub}
\usepackage{mathtools, amsfonts, amsthm, latexsym, amssymb, dsfont, mathrsfs}
\usepackage[T1]{fontenc}
\usepackage{newtxtext}
\usepackage{physics2}
\usephysicsmodule{ab}
\usepackage{braket}
\usepackage{color}
\usepackage{bm}
\usepackage{comment}

\usepackage{tikz}
\usetikzlibrary{decorations.markings}

\usepackage{hyperref}
\hypersetup{colorlinks=true, citecolor=blue, urlcolor=blue}
\usepackage{cleveref}

\usepackage{titlesec}
\titlespacing{\section}
    {0pt}% left
    {5pt}% before-sep
    {3pt}% after-sep
\titlespacing{\subsection}
    {0pt}% left
    {5pt}% before-sep
    {3pt}% after-sep

\AtBeginDocument{
  \abovedisplayskip     =0.7\abovedisplayskip
  \abovedisplayshortskip=0.7\abovedisplayshortskip
  \belowdisplayskip     =0.7\belowdisplayskip
  \belowdisplayshortskip=0.7\belowdisplayshortskip}


\newcommand{\del}{\partial}

\begin{document}

\title{Revisiting boundary electromagnetic duality\\ and edge modes}

\author[a]{Keito Shimizu}
\author[b, c]{and Sotaro Sugishita}

\affiliation[a]{Department of Physics, Kyoto University, Kitashirakawa-Oiwakecho, Kyoto 606-8502, Japan}
\affiliation[b]{Department of Physics, Hokkaido University, Sapporo 060-0810, Japan}
\affiliation[c]{RIKEN Center for Interdisciplinary Theoretical and Mathematical Sciences (iTHEMS), RIKEN, Wako 351-0198, Japan}
\date{}

\usetikzlibrary{calc} 

\emailAdd{kate@gauge.scphys.kyoto-u.ac.jp, sugishita(at)particle.sci.hokudai.ac.jp}

\abstract{
    We revisit electric and magnetic surface charges and edge modes in four-dimensional Maxwell theory and QED on a spacetime with a finite spatial boundary. 
    Using the S-wall, which implements electromagnetic duality, we clarify the dual structure of surface charges. 
    We show that, for the standard Neumann and Dirichlet boundary conditions, large gauge transformations and the corresponding shifts of edge modes are gauge redundancies rather than physical boundary symmetries. 
    We also consider singular large gauge transformations and interpret them as insertions of Wilson or 't Hooft loops on the boundary. 
    For modified boundary conditions, we show that large gauge transformations can become genuine physical boundary symmetries generated by topological surface operators, and that the corresponding edge modes can become physical.
    We further construct new boundary conditions that are electromagnetic duals of the modified boundary conditions.}

\preprint{%
\begin{tabular}[t]{@{}r@{}}
KUNS-3105\\
EPHOU-26-05\\
RIKEN-iTHEMS-Report-26
\end{tabular}%
}

\maketitle
\newpage

\section{Introduction}
Gauge theory has been a central subject in physics, and its rich and intricate symmetry structure has attracted many physicists.
In order to explore it, some physicists have investigated symmetry itself, and others have placed gauge theory on various stages --- namely, spacetime with boundaries at finite/infinite distance.

The presence of a boundary exposes a fascinating aspect of gauge theories. 
Gauge transformations that generate gauge redundancies in the bulk can be associated with nontrivial physical charges or boundary degrees of freedom (edge modes) once we consider a boundary or an asymptotic region. 
The importance of boundary degrees of freedom has been recognized for a long time. 
In the Hamiltonian formulation of general relativity, Regge and Teitelboim \cite{Regge:1974otg, Regge:1974zd} showed that suitable surface terms are required for the Hamiltonian generators and these terms give the conserved charges on the physical phase space. 
Related ideas were developed in gauge theories by Gervais, Sakita and Wadia, who emphasized the role of surface terms and surface variables \cite{Gervais:1976ec} (see also \cite{Wadia:1976fa, Wadia:1977qr}).
Surface charges also play an important role in the work by Brown and Henneaux demonstrating the central extension of the asymptotic symmetry algebra in three-dimensional gravity in asymptotically AdS space \cite{Brown:1986nw}. 
Another interesting example is the Chern--Simons theory on a manifold with boundary, where degrees of freedom that would be gauge in the bulk appear as physical boundary modes \cite{Witten:1988hf, Bos:1989kn, Elitzur:1989nr, Balachandran:1991dw}. 
More generally, gauge theories on manifolds with boundaries are studied and it is emphasized that boundary-localized observables can arise in various gauge theories, with their existence depending on the imposed boundary conditions \cite{Balachandran:1994vi}.
A similar analysis is also done for gravity in \cite{Balachandran:1994up}, where it is argued that removing a spatial region leads to associated edge states, and their possible relevance to black hole physics is discussed. 
More recent studies have also emphasized that edge modes provide a useful framework for describing various aspects of gauge theories, such as the entanglement entropy \cite{Donnelly:2014fua, Donnelly:2015hxa, David:2022jfd, Ball:2024hqe} and the definition of gauge invariant subsystems \cite{Donnelly:2016auv, Speranza:2017gxd}.

Recently, surface charges and edge modes in gauge theories have been intensively studied through asymptotic symmetry on asymptotically flat spacetime.
The symmetry is closely related to infrared properties of gauge theories and gravity such as soft theorems, memory effects, dressed states \cite{Balachandran:2013wsa, Strominger:2013lka, Strominger:2013jfa, He:2014laa, Cachazo:2014fwa, Campiglia:2014yka, He:2014cra, Kapec:2014opa, Lysov:2014csa, Liu:2014vva, Strominger:2014pwa, Kapec:2014zla, Larkoski:2014bxa, Kapec:2015vwa, He:2015zea, Campiglia:2015qka, Kapec:2015ena, Campiglia:2015kxa, Pasterski:2015tva, Dumitrescu:2015fej, Strominger:2015bla, Campiglia:2016hvg, Mirbabayi:2016axw, Gabai:2016kuf, Campiglia:2017dpg, Kapec:2017tkm, Hamada:2017atr, Pate:2017fgt, Hamada:2018cjj, Carney:2018ygh, Hirai:2018ijc, 
Hosseinzadeh:2018dkh, Campiglia:2018see, Francia:2018jtb, Neuenfeld:2018fdw, Henneaux:2018mgn, Hirai:2019gio, Gonzo:2019fai, Choi:2019rlz, Choi:2019sjs, Henneaux:2020nxi, Hirai:2020kzx, Campiglia:2021oqz, Hirai:2022yqw, Nagy:2022xxs, Peraza:2023ivy, Nagy:2024jua, Oertel:2026wsm, Oertel:2026oqv} and confinement \cite{Shimizu:2025hfl}.
In addition, this symmetry is relevant to the analysis of holographic aspects of gravity through celestial holography \cite{Pasterski:2016qvg, Pasterski:2017ylz, Cardona:2017keg, Pasterski:2017kqt, Arkani-Hamed:2020gyp, Donnay:2020guq, Raclariu:2021zjz, Pasterski:2021rjz, Iacobacci:2022yjo, Sleight:2023ojm, Hao:2023wln, Furugori:2023hgv, Iacobacci:2024nhw, Banerjee:2024hvb, Melton:2024akx, Furugori:2025xkl, Furugori:2025lgu}.

On the other hand, according to recent generalizations of symmetry \cite{Gaiotto:2014kfa}, symmetry operators should be expressed as topological operators.
However, asymptotic symmetries are not yet fully understood in terms of topological operators, although some studies have tried to connect these two notions of symmetry, e.g. \cite{Lake:2018dqm, Berean-Dutcher:2025ohp, Tizzano:2026rgr}.
Consequently, there are many features of gauge theory which are known only in terms of asymptotic symmetry or canonical formalism, such as a dual relation of electric and magnetic asymptotic charges in Maxwell theory and quantum electrodynamics (QED), which was first suggested \cite{Strominger:2015bla} in the magnetic picture with monopoles and soon after found \cite{Campiglia:2016hvg} in the electric picture without monopoles\footnote{
Although this may seem confusing at first, in the former work, monopoles are described in the dual picture, which is equivalent to describing electrically charged matter in ordinary electromagnetism.
}, followed by several studies 
\cite{Hosseinzadeh:2018dkh, Freidel:2018fsk, Choi:2019sjs, Mathieu:2020fwg, Henneaux:2020nxi,  Geiller:2021gdk}. See also \cite{Balachandran:2021sze}, where the magnetic counterpart of large gauge charges is studied.

Electromagnetic duality is originally an invariance of the Maxwell equation under the exchange of electric field $\bm{E}$ and magnetic field $\bm{B}$.
This can also be understood as a duality between the equation of motion (EoM) $d*F=0$ and the Bianchi identity $dF=0$ under exchange of $*F$ and $F$.
The presence of electrically charged matter breaks this duality explicitly unless magnetically charged matter is introduced since electrically charged matter appears on the right-hand side of the EoM as $d*F=*J$, while the Bianchi identity still holds.
Nevertheless, it is suggested in \cite{Campiglia:2016hvg} that we can define magnetic asymptotic charges, which seem to be dual to electric asymptotic charges,  even in the absence of monopoles.
In \cite{Hosseinzadeh:2018dkh,Freidel:2018fsk}, the magnetic charges are interpreted as charges corresponding to large gauge symmetry (or surface symmetry of magnetic edge mode) of the dual gauge potential.
Although it is impossible to define the dual gauge potential in the bulk in the presence of electrically charged matter, which plays the same role as monopoles in QED, it can be defined, in a certain sense, at the boundary if charged matter cannot reach the boundary.

Our main purpose in this paper is to revisit electric and magnetic surface charges in Maxwell theory and QED on a spacetime with a finite spatial boundary from the perspective of generalized symmetries, and to clarify the meaning of these charges and their duality.
For this purpose, we utilize the S-wall \cite{Kapustin:2009av}, which is a topological operator implementing the electromagnetic duality.
The S-wall is helpful to see the electromagnetic dual structure of surface charges and to clarify the meaning of boundary electromagnetic duality explored in \cite{Freidel:2018fsk}.
Using the S-wall, we also find new boundary conditions dual to those studied in \cite{Ball:2024hqe, Araujo-Regado:2024dpr}.
We argue that in the standard Neumann or Dirichlet condition, the electric and magnetic surface charges generate gauge redundancies and thus these charges are not physical. 
For the modified boundary conditions studied in \cite{Ball:2024hqe, Araujo-Regado:2024dpr}, we can construct physical electric surface charges, while for the dual new boundary conditions, we can construct magnetic ones.
We further point out that we can regard these surface charges as topological operators.
This may enable us to interpret asymptotic symmetries and surface symmetries as boundary higher form symmetries, which have been studied in the context of symmetry topological field theory (see \cite{Arbalestrier:2025jsg} for a recent work studying the relationship between boundary conditions and symmetry in Maxwell theory comprehensively).

The paper is organized as follows.
In \cref{sec:warm-up}, we give a brief %lightning
review of generalized symmetry by taking Maxwell theory as an example.
In \cref{sec:elemag dual}, we consider Maxwell theory on a spacetime with a boundary, and study the surface charges generating the large gauge transformations (or shift of edge modes) with the standard Neumann and Dirichlet conditions. 
We show that in the Neumann condition, we can write down the magnetic surface charges dual to the electric ones in the Dirichlet condition by introducing a hidden magnetic edge mode. 
We also argue that these charges generate gauge redundancies and discuss the singular transformations. 
We also consider generalization to QED, and clarify the boundary electromagnetic duality.
In \cref{sec:Retake}, we re-examine boundary conditions to find non-trivial symmetry transformations and physical surface charges.
We also find dual counterparts to some known boundary conditions.
Finally, we conclude the paper in \cref{sec:summary} with a summary and discussion.

%%%%%%%%%%%%%%%%%%%
\section{Warm-up: symmetry as topological operator}\label{sec:warm-up}
First of all, let us briefly review the relationship between generalized symmetry and topological operator \cite{Gaiotto:2014kfa} through the four-dimensional Maxwell theory
\begin{align}\label{action:Maxwell}
    S[A] = -\frac{1}{2e^2}\int F\wedge*F,\quad F=dA.
\end{align}
For the time being, we do not care whether the spacetime has a boundary since it does not affect the main point of the discussion.

Maxwell theory has a shift symmetry of the gauge field
\begin{align}\label{eq:electric 1-form symmetry}
    A\to A+\lambda,
\end{align}
where $\lambda$ is a one-form satisfying $d\lambda=0$.
According to Noether's theorem, there is a conserved 2-form current $*J$ corresponding to the symmetry as
\begin{align}
    *J = *F.
\end{align}
It is indeed conserved on-shell $d*J=d*F~\hat{=}~0$, where $\hat{=}$ denotes on-shell evaluation.
This current allows us to construct a conserved charge associated with a closed two-dimensional surface
\begin{align}
    Q^{(1)}_e[\Sigma_2] = \frac{1}{e^2}\int_{\Sigma_2} *F.
\end{align}
\begin{figure}
    \centering
    \includegraphics[width=0.3\linewidth]{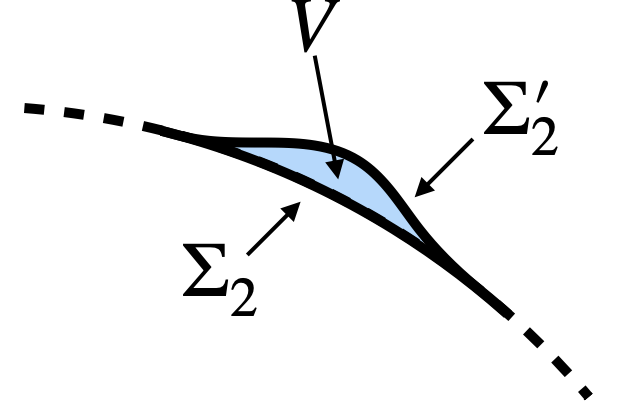}
    \caption{
    A two-dimensional surface $\Sigma_2'$ is obtained by deforming $\Sigma_2$.
    $V$ is a three dimensional region surrounded by $\Sigma_2$ and $\Sigma_2'$.}
    \label{fig:topologicalness}
\end{figure}
This conservation law is reflected in the topological property of the corresponding charge.
Indeed, the charge is independent of the surface $\Sigma_2$ since it is invariant under deformations (see \cref{fig:topologicalness})
\begin{align}
    Q^{(1)}_e[\Sigma_2'] - Q^{(1)}_e[\Sigma_2]
    &= \frac{1}{e^2}\int_{\Sigma_2'-\Sigma_2}*F = \frac{1}{e^2}\int_{V~(\del V=\Sigma_2'-\Sigma_2)}d*F = 0.
\end{align}
Although it has been assumed that the charge is associated with closed surfaces, the topological feature still holds for surfaces with boundaries as long as boundaries are not changed under the deformation.
Analogously, one can construct another conserved charge as
\begin{align}
    Q^{(1)}_m[\Sigma_2] = \int_{\Sigma_2}F,
\end{align}
where the 2-form current $F$ is also conserved by the Bianchi identity $dF=0$. Thus, the charge $Q^{(1)}_m[\Sigma_2]$ is also topological.
This charge is associated with a shift symmetry of the dual gauge potential.
The first symmetry generated by $Q^{(1)}_e$ is called the electric $U(1)$ 1-form symmetry, while the second one by $Q^{(1)}_m$ is called the magnetic $U(1)$ 1-form symmetry.

In summary, if the theory has a continuous symmetry, one has a conserved current according to Noether's theorem, and one can construct a topological operator by integrating a conserved current on a suitable submanifold.
Actually, one can construct a topological operator that implements a symmetry transformation even in the absence of conserved currents, as in the case of discrete symmetries.
In a modern view, symmetry is understood as the existence of an associated topological operator.

\section{Electromagnetic dual structure at boundary}\label{sec:elemag dual}
In this section, we begin by introducing a topological operator called the S-wall \cite{Kapustin:2009av}, which implements the electromagnetic duality, and then we see the duality transformation between surface charges of Maxwell theory by means of the S-wall.
Next, we introduce edge modes in the Neumann boundary condition to find a missing counterpart of electric surface charges under the Dirichlet boundary condition.
We also consider surface charges for singular transformations.
Finally, we comment on the case with charged matter, boundary electromagnetic duality, and existing works.

We will consider theories in a (3+1)-dimensional flat spacetime $\mathcal{M}$ with a (2+1)-dimensional boundary $\Delta$.
Although the specific shape of the boundary is not important for the following discussion, for concreteness we take $\mathcal{M}$ to be a solid cylinder with boundary $\Delta$ at $r=R$.
A metric of the bulk is given by $ds^2=-dt^2+dr^2 + r^2\gamma_{ab}d\Omega^ad\Omega^b$ with angular variables $\Omega^a~(a=1,2)$ and $\gamma$ is the metric for the unit sphere $S^2$.
Each volume form is given by $\mathrm{vol}_{\mathcal{M}}=\frac{1}{2}\varepsilon_{ab}\sqrt{-g}~dt\wedge dr\wedge d\Omega^a\wedge d\Omega^b$ and $\mathrm{vol}_{\Delta}=\frac{1}{2}\varepsilon_{ab}R^2\sqrt{\gamma}~dt\wedge d\Omega^a\wedge d\Omega^b$ respectively, where $\varepsilon^{12}=-\varepsilon^{21}=\varepsilon_{12}=-\varepsilon_{21}=1$ and $\varepsilon^{11}=\varepsilon^{22}=\varepsilon_{11}=\varepsilon_{22}=0$.
These volume forms fix orientations of the bulk and the boundary.

\subsection{Review of S-wall}
Maxwell theory is self-dual under the electromagnetic duality, and it implies that the theory has a topological operator corresponding to this symmetry. 
It is the S-wall.
To see how the S-wall works, let us first consider Maxwell theory on a spacetime $\mathcal{M}'$ without boundaries, 
\begin{align}
    \int\mathcal{D}A~e^{iS[A]} = \int\mathcal{D}A~e^{-\frac{i}{2e^2}\int_{\mathcal{M}'}F\wedge*F}.
\end{align}
The S-wall is a wall operator $W[A_+,A_-]$ defined on a three dimensional surface $W$ as
\begin{align}
    \int\mathcal{D}A_+\mathcal{D}A_-~e^{iW[A_+,A_-]}e^{iS_+[A_+]+iS_-[A_-]}
\end{align}
with the Maxwell action on two sides $\mathcal{M}'_\pm$ divided by $W$
\begin{align}
    S_+[A_+]+S_-[A_-] = -\frac{1}{2e_+^2}\int_{\mathcal{M}'_+}F_+\wedge*F_+  -\frac{1}{2e_-^2}\int_{\mathcal{M}'_-}F_-\wedge*F_-
\end{align}
and
\begin{align}
    W[A_+,A_-] = \frac{1}{2\pi}\int_W A_+\wedge dA_-.
\end{align}
The variational principle requires
\begin{align}
    \frac{1}{e_+^2}*F_+ \overset{W}{=} -\frac{1}{2\pi}F_-,\quad \frac{1}{e_-^2}*F_- \overset{W}{=} \frac{1}{2\pi}F_+,
\end{align}
where $\overset{W}{=}$ represents the equality restricted on $W$ while $\ast$ is the hodge star on four-dimensional spacetime.
In addition, the topologicality of the S-wall requires
\begin{align}
    e_+ = \frac{2\pi}{e_-}.
\end{align}
These equations precisely represent the electromagnetic-duality relations in Maxwell theory.

We here summarize the terminology of the boundary conditions before considering the S-wall in the spacetime $\mathcal{M}$ with the boundary $\Delta$.
In this paper, the Neumann boundary condition means that the restriction of the two-form $\ast F$ onto the boundary $\Delta$ is fixed, \textit{i.e.}, 
\begin{align}
    \text{Neumann: } \ast F \overset{\Delta}{=} \text{fixed},
\end{align}
where $\overset{\Delta}{=}$ represents the equality on $\Delta$.
The Dirichlet boundary condition means that 
$F$ is fixed at $\Delta$, \textit{i.e.}, 
\begin{align}
    \text{Dirichlet: } F \overset{\Delta}{=} \text{fixed}.
\end{align}

We now consider the S-wall in the presence of boundary $\Delta$.
Since the electromagnetic duality exchanges $F$ and $*F$, the Neumann boundary condition should be replaced by the Dirichlet boundary condition and vice versa after dualiy.
\begin{figure}
    \centering
    \includegraphics[width=0.3\linewidth]{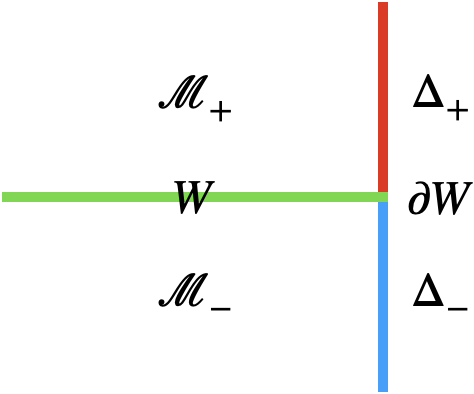}
    \caption{
    The spacetime is separated into two parts $\mathcal{M}_\pm$ with boundary $\Delta_\pm$ by a wall $W$.
    The Dirichlet boundary condition is satisfied on $\Delta_+$ and the Neumann boundary condition is satisfied on $\Delta_-$, respectively.
    }
    \label{fig:wall operator}
\end{figure}
One can understand this situation precisely by inserting the S-wall attached to the boundary $\Delta$ and assuming the Dirichlet condition and Neumann condition are realized on $\Delta_\pm$ respectively, see \cref{fig:wall operator}.
The theory describing this setup is given by
\begin{equation}
    \begin{gathered}
        \int\mathcal{D}\mu~e^{iW[A_+,A_-]}e^{iS_+[A_+,\hat{A};\mathcal{A}]+iS_-[A_-;\mathcal{A}]},\quad
        \mathcal{D}\mu\equiv\mathcal{D}A_+
        \mathcal{D}A_-\mathcal{D}\hat{A},
    \end{gathered}
\end{equation}
with
\begin{align}
    S_-[A_-;\mathcal{A}] &= -\frac{1}{2e_-^2}\int_{\mathcal{M}_-}F_-\wedge*F_- - \frac{1}{2\pi}\int_{\Delta_-} A_-\wedge d\mathcal{A},\label{eq:S_-} \\
    S_+[A_+,\hat{A};\mathcal{A}] &= -\frac{1}{2e_+^2}\int_{\mathcal{M}_+}F_+\wedge*F_+ - \frac{1}{2\pi}\int_{\Delta_+} (A_+\wedge d\hat{A}-\hat{A}\wedge d\mathcal{A}),\label{eq:S_+}
\end{align}
where $\mathcal{A}$ is a non-dynamical and fixed background gauge field, and $\hat{A}$ is an auxiliary gauge field introduced on the boundary. We soon verify that this boundary auxiliary leads to the Dirichlet boundary condition below. 

To ensure a well-posed variational principle, we demand that the variation of the action vanish at $\Delta_\pm$, $W$ and the corner $\del W$ for arbitrary variations.
We then obtain
\begin{equation}
    \begin{gathered}
        \frac{1}{e_+^2}*F_+ \overset{W}{=} -\frac{1}{2\pi}F_-,\quad \frac{1}{e_-^2}*F_- \overset{W}{=} \frac{1}{2\pi}F_+, \\
        \frac{1}{e^2_-}*F_-\overset{\Delta_-}{=}\frac{1}{2\pi}d\mathcal{A},\quad \frac{1}{e^2_+}*F_+\overset{\Delta_+}{=}\frac{1}{2\pi}d\hat{A},\quad F_+\overset{\Delta_+}{=}d \mathcal{A}.\label{eq:boundary conditions with the S-wall}
    \end{gathered}
\end{equation}
They show that the Neumann boundary condition, $\ast F = \text{fixed}$, is imposed on $\Delta_-$ and the Dirichlet one, $F = \text{fixed}$, is imposed on $\Delta_+$.
From the variation on $\del W$, we
also obtain
\begin{align}\label{eq:bdy cond at corner}
    \frac{1}{2\pi}\int_{\del W}A_+\wedge(\delta A_-+\delta\hat A) = 0,
\end{align}
where the first term comes from the S-wall while the second term from $S_+$.
The vanishing condition \eqref{eq:bdy cond at corner} requires
\begin{align}\label{A-=-hatA}
    A_-\overset{\del W}{=}-\hat A.
\end{align}

\begin{figure}
    \centering
    \includegraphics[width=0.3\linewidth]{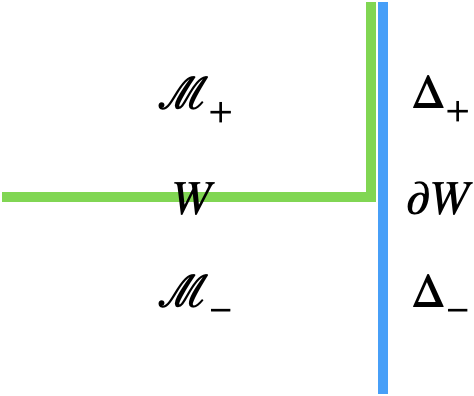}
    \caption{
    The bending S-wall operator is inserted in the spacetime.
    It separates the spacetime into two parts by $W$ and goes along the boundary $\Delta_+$ to the future time-like infinity.
    }
    \label{fig:bending wall operator}
\end{figure}

This result and the form of \eqref{eq:S_+} imply that we should interpret this setup as a system with a bending S-wall as depicted in \cref{fig:bending wall operator}, not with a cut S-wall attached to the boundary.
The interpretation also makes the gauge invariance and the topologicality manifest.
One can find similar descriptions of symmetry defects in a spacetime with boundaries in \cite{Choi:2023xjw}.

\subsection{Charges}

Asymptotic charges are associated with the ``large gauge transformations'' in the asymptotic region. 
We consider their counterparts in the case of the finite boundary. 
It is known that there are two types of soft charges (electric and magnetic) \cite{Strominger:2015bla, Campiglia:2016hvg, Hosseinzadeh:2018dkh, Freidel:2018fsk, Choi:2019sjs, Mathieu:2020fwg, Henneaux:2020nxi, Geiller:2021gdk}, as mentioned in the introduction.
Since the S-wall exchanges the electric field and magnetic field, it is expected that the electric and magnetic asymptotic charges are also exchanged.\footnote{We use the term \textit{asymptotic charges} even in the finite-boundary case.}
With this in mind, let us consider charges generating large gauge transformations under the Neumann and Dirichlet boundary conditions.
For simplicity, we just consider the theory under the Neumann boundary condition or the Dirichlet boundary condition, not a theory with the two conditions connected by the S-wall considered in the previous subsection.

We first consider the Neumann boundary condition. 
The action is $S_-[A_-; \mathcal{A}]$ given by \eqref{eq:S_-}.
This action has the large gauge symmetry 
\begin{align}\label{eq:large gauge symmetry under N}
    A_-\to A_-+d\alpha_-,
\end{align}
where $\alpha_-$ is an arbitrary gauge parameter which does not vanish at the boundary $\Delta_-$.
A naive charge generating the transformation is given by 
\begin{align}\label{eq:charge Q_e^-}
    Q_e^-[\alpha_-]&=-\frac{1}{e_-^2}\int_{\Sigma_2(\subset\Delta_-)}\alpha_-\ast F_-,
\end{align}
where $\Sigma_2$ is a two-dimensional spacelike surface on the boundary.
This expression of charge is the same as the electric asymptotic charges.
Using the Neumann boundary condition $\frac{1}{e_-^2}*F\overset{\Delta_-}{=}\frac{1}{2\pi}d\mathcal{A}$ in \eqref{eq:boundary conditions with the S-wall}, we can rewrite it as
\begin{align}\label{eq:charge Q_e^-2}
    Q_e^-[\alpha_-]&=-\frac{1}{2\pi}\int_{\Sigma_2(\subset\Delta_-)}\alpha_-d\mathcal{A}.
\end{align}

However, \eqref{eq:charge Q_e^-2} is just a number because $\mathcal{A}$ is a background field introduced by hand. 
It means that $Q_e^-[\alpha_-]$ generates nothing on the physical phase space.
Indeed, the large gauge transformation \eqref{eq:large gauge symmetry under N} is degenerate with respect to the \mbox{(pre-)symplectic} form (see Appendix~\ref{sec:symplectic form and charge} for detailed calculations).\footnote{\eqref{eq:charge Q_e^-} generates the transformation \eqref{eq:large gauge symmetry under N} on an extended phase space without imposing the boundary EoM $\frac{1}{e_-^2}*F\overset{\Delta_-}{=}\frac{1}{2\pi}d\mathcal{A}$. This is consistent with the fact that \eqref{eq:charge Q_e^-2} does not generate any transformation, because \eqref{eq:large gauge symmetry under N} is a gauge redundant direction after imposing the condition $\frac{1}{e_-^2}*F\overset{\Delta_-}{=}\frac{1}{2\pi}d\mathcal{A}$. In Appendix~\ref{sec:symplectic form and charge}, we show that the Noether current explicitly vanishes and thus the corresponding charge vanishes.}
Thus, the large gauge transformation \eqref{eq:large gauge symmetry under N} is not a physical symmetry but a gauge redundancy under the Neumann boundary condition.

We next consider the Dirichlet boundary condition. 
The action is $S_+[A_+,\hat{A};\mathcal{A}]$ given by \eqref{eq:S_+}.
We can consider two large gauge symmetries 
\begin{align}\label{eq:large gauge symmetry under D}
    A_+\to A_++d\alpha_+,\quad \hat A\to \hat A+d\hat\alpha.
\end{align}
As in the Neumann case, naive charges generating these transformations are given by (see the Appendix~\ref{sec:symplectic form and charge} for details)
\begin{align}
    Q_e^+[\alpha_+]&=-\frac{1}{2\pi}\int_{\Sigma_2(\subset\Delta_+)}d(\alpha_+\hat{A}),\label{eq:charge Q_e^+}\\
    Q_m^+[\hat\alpha]&=\frac{1}{2\pi}\int_{\Sigma_2(\subset\Delta_+)} A_+\wedge d\hat\alpha.
    \label{eq:charge Q_m^+}
\end{align}
We call the first one the electric gauge charge because it is associated with the large gauge transformation of the bulk gauge field $A_+$, similar to the Neumann case.
We call the second one the magnetic gauge charge, which can be found in \cite{Freidel:2018fsk}.

As in the Neumann case, these charges are not physical in the sense that they generate just gauge redundancy as follows.
First, the integrand of $Q_e^+$ in \eqref{eq:charge Q_e^+} is always a total derivative. Thus, if we consider a globally well-defined parameter $\alpha_+$, this charge always vanishes. Thus, the associated transformation $A_+ \to A_+ + d\alpha_+$  must be a gauge redundancy, not a physical symmetry. 
Next, $Q_m^+$ given by \eqref{eq:charge Q_m^+} can be written as
\begin{align}
     Q_m^+[\hat\alpha]&=\frac{1}{2\pi}\int_{\Sigma_2(\subset\Delta_+)} \hat\alpha F_+=\frac{1}{2\pi}\int_{\Sigma_2(\subset\Delta_+)} \hat\alpha d \mathcal{A},
     \label{eq:charge Q_m^+2}
\end{align}
where we have used the Dirichlet boundary condition $F_+\overset{\Delta_+}{=}d \mathcal{A}$.
Thus, this charge $Q_m^+$ is just a number like $Q_e^-$ in the Neumann case, and it does not generate a physical symmetry.

To summarize the discussion so far, for the standard Neumann and Dirichlet boundary conditions, all large gauge transformations are gauge redundancies and not physical symmetries. In order to make large gauge transformations at the boundary into physical symmetries, one has to impose different boundary conditions. We will consider such boundary conditions in \cref{sec:Retake}.

Nevertheless, the expression for a charge that generates a gauge transformation is not completely meaningless. For example, such a charge can be useful when one discusses quantization, as in the BRST quantization. 
Thus, let us consider the duality relations for these gauge charges.

The magnetic charge $Q_m^+$ \eqref{eq:charge Q_m^+2} for the Dirichlet condition is dual to the electric one $Q_e^-$ \eqref{eq:charge Q_e^-2} for the Neumann condition. 
Indeed, one can find that $Q_m^+$ is mapped to $Q_e^-$ under the identification $\hat{\alpha}=-\alpha_-$, which is consistent with \eqref{A-=-hatA}.
Thus, the large gauge transformation $\hat A\to \hat A+d\hat\alpha$ in the Dirichlet case is dual to the one $A_-\to A_-+d\alpha_-$ in the Neumann case.

It seems that there is no dual pair of $A_+\to A_++d\alpha_+$.
This is not a problem because this symmetry is just a gauge transformation, and the corresponding charge $Q_e^+$ \eqref{eq:charge Q_e^+} always vanishes. 
The two dual theories have the same physical degrees of freedom after quotienting out the gauge redundancy.

One still might think that we should have a magnetic gauge charge $Q_m^-$ for the Neumann boundary condition such that it is the dual expression of $Q_e^+[\alpha_+]$, because $Q_e^+[\alpha_+]$ does not vanish if we consider a singular gauge parameter $\alpha_+$ as we will see in \cref{sec:singular}.
If we have such a $Q_m^-$, the duality becomes more explicit including the gauge redundancy.  
As we explain below, this can be achieved by introducing edge modes.

\subsection{Adding edge modes}
We now introduce edge modes \cite{Donnelly:2016auv} in the system.
After introducing them, we derive a modified \mbox{(pre-)symplectic} form where the duality of the gauge charges is more explicit.
In other words, we can write down a magnetic gauge charge in the Neumann boundary condition dual to the electric charge $Q_e^+[\alpha_+]$ in the Dirichlet one.

Edge modes can be added by dressing up the gauge field on the boundary with a scalar (gauge) field $\varphi$ as\footnote{This edge mode is introduced by hand to have a gauge-invariant combination \cite{Donnelly:2016auv}, and it is a kind of the Stueckelberg field living on the boundary.
A systematic procedure to obtain a gauge-invariant combination is studied in \cite{Fournel:2012cr, Francois:2024rdm}
See also \cite{Nagy:2025hip}.
}
\begin{align}
    a \equiv A|_\Delta + d\varphi,
\end{align}
which gives the same field strength $da=dA=F$.
This dressed gauge field $a$ is invariant under the gauge transformation $(A,\varphi)\to(A+d\alpha,\varphi-\alpha)$. 
In this picture, the action with the Neumann boundary condition is given by
\begin{align}\label{eq:action under N}
    S_N[A,\varphi;\mathcal{A}]
    = -\frac{1}{2e^2}\int_{\mathcal{M}}F\wedge*F - \frac{1}{2\pi}\int_\Delta a\wedge d\mathcal{A}.
\end{align}
For the Dirichlet boundary condition, we also introduce the edge mode $\hat{\varphi}$ for $\hat{A}$ as 
\begin{align}
    \hat{a} \equiv \hat{A} + d\hat\varphi.
\end{align}
The action with the edge modes in the Dirichlet condition is given by 
\begin{align}\label{eq:action under D}
    &S_D[A,\hat{A},\varphi,\hat\varphi;\mathcal{A}]= -\frac{1}{2e^2}\int_{\mathcal{M}}F\wedge*F - \frac{1}{2\pi}\int_{\Delta} (a\wedge d\hat{a}-\hat{a}\wedge d\mathcal{A}).
\end{align}

With the edge modes, the actions are invariant under the gauge transformation $(A,\varphi)\to(A+d\alpha,\varphi-\alpha)$, $(\hat{A},\hat{\varphi})\to(\hat{A}+d\hat{\alpha},\hat{\varphi}-\hat{\alpha})$, and also invariant under the shift transformation of the edge modes $(A,\varphi)\to(A,\varphi+\alpha)$ and $(\hat{A},\hat{\varphi})\to(\hat{A},\hat{\varphi}+\hat{\alpha})$ \cite{Donnelly:2016auv}.
The \mbox{(pre-)symplectic} forms and the charge corresponding to these transformations are summarized in Appendix~\ref{sec:symplectic form and charge}. 
One can find the dual relation between $\varphi \to \varphi +\alpha$ in $S_N$ and $\hat\varphi \to \hat\varphi +\hat\alpha$ in $S_D$ as in the previous subsection.

The (gauge) charge for $\varphi \to \varphi +\alpha$ in $S_D$ is given by
\begin{align}
    Q_e^D[\alpha]=-\frac{1}{2\pi}\int_{\del\Sigma}d(\alpha\hat a).
    \label{Q_e^Dalpha}
\end{align}
What we want to find is the dual counterpart of this transformation, $\varphi \to \varphi+\alpha$, or equivalently, the dual counterpart of this gauge charge $Q^D_e[\alpha]$.
This can be realized by further introducing another edge mode to \eqref{eq:action under N} using the ambiguity of the symplectic form.
The ambiguity is as follows. 
Variation of the action \eqref{eq:action under N} gives
\begin{align}\label{deltaS_N_again}
    \delta S_N
    &= -\frac{1}{e^2}\int_{\mathcal{M}}\delta A\wedge d*F  -\frac{1}{e^2}\int_{\mathcal{M}}d(\delta A\wedge*F) - \frac{1}{2\pi}\int_{\Delta}\delta a\wedge d\mathcal{A}.
\end{align}
According to the prescription proposed in \cite{Harlow:2019yfa}, it is required that
\begin{align}
    \ab(\frac{1}{e^2}\delta A\wedge*F)\biggr|_{\Delta} - \frac{1}{2\pi}\delta a\wedge d\mathcal{A}=dc
\end{align}
at the boundary\footnote{Be careful about the sign of the first term due to $\mathrm{vol}_{\mathcal{M}}=\frac{1}{2}\varepsilon_{ab}\sqrt{-g}~dt\wedge dr\wedge d\Omega^a\wedge d\Omega^b=-\frac{1}{2}\varepsilon_{ab}\sqrt{-g}~dr\wedge dt\wedge d\Omega^a\wedge d\Omega^b$. We have chosen the orientation of $\Delta$ as $\mathrm{vol}_{\Delta}=\frac{1}{2}\varepsilon_{ab}R\sqrt{-\gamma}~dt\wedge d\Omega^a\wedge d\Omega^b$.} where $dc$ is an exact form.
Thus, the \mbox{(pre-)symplectic} form on a Cauchy slice $\Sigma$ is given by
\begin{align}
    \Omega_{N,\Sigma} = \frac{1}{e^2}\int_{\Sigma}\delta A\wedge\delta*F +\int_{\del\Sigma} \delta c.
\end{align}
Imposing the Neumann boundary condition $\frac{1}{e^2}*F\overset{\Delta}{=}\frac{1}{2\pi}d\mathcal{A}$, one may take $c=\frac{1}{2\pi}\delta\varphi d\mathcal{A}$, which does not contribute to the \mbox{(pre-)symplectic} form because the boundary field $\mathcal{A}$ is not dynamical.
However, there is a redundancy $c\to c+dY$ by construction.
We take $dY$ to be 
\begin{align}
    dY=\frac{1}{2\pi}d(\tilde{\varphi}\cdot\delta a),
\end{align}
where $\tilde\varphi$ is a hidden edge mode.
This addition modifies $\Omega_{N,\Sigma}$ as\footnote{The term $\delta\varphi\cdot\delta*F$ is vanishing and is just a formal expression because $\ast F$ is fixed at the boundary $\frac{1}{e^2}*F\overset{\Delta}{=}\frac{1}{2\pi}d\mathcal{A}$.}
\begin{align}\label{eq:modified symplectic form}
    \Omega_{N,\Sigma}
    &= \frac{1}{e^2}\int_{\Sigma}\delta A\wedge\delta*F + \frac{1}{2\pi}\int_{\del\Sigma} \ab(\frac{2\pi}{e^2}\delta\varphi\cdot\delta*F+d(\delta\tilde{\varphi}\cdot\delta a)).
\end{align}
It has a gauge symmetry
\begin{align}\label{eq:hidden magnetic symmetry}
    \tilde\varphi\to\tilde\varphi+\tilde\alpha
\end{align}
with the gauge charge%\footnote{This charge seems to vanish as well as \eqref{eq:charge Q_e^+}. We also discuss it in \cref{sec:Retake}.}
\begin{align}\label{eq:charge Q_m^-}
    Q_m^N[\tilde\alpha] = -\frac{1}{2\pi}\int_{\partial \Sigma}d(\tilde\alpha a).
\end{align}
Noting the dual relation $a=-\hat{a}$ as \eqref{A-=-hatA} where $a$ on the left-hand side represents the dressed gauge field for the Neumann case, we can find that this $Q_m^N[\tilde\alpha]$ is the very dual counterpart of $Q_e^D[\alpha]$ in \eqref{Q_e^Dalpha} by identifying $\tilde{\alpha}=-\alpha$.

The modification \eqref{eq:modified symplectic form} corresponds to adding a total derivative term to the original action $S_N$ as 
\begin{align}\label{eq:action under N with mag edge}
    S=-\frac{1}{2e^2}\int_{\mathcal{M}}F\wedge*F - \frac{1}{2\pi}\int_{\Delta}a\wedge d\mathcal{A} + \frac{1}{2\pi}\int_{\Delta}da\wedge d\tilde\varphi.
\end{align}
The added field $\tilde\varphi$ is the magnetic edge mode studied in \cite{Freidel:2018fsk}, and it can be a dynamical variable in this sense even though $\mathcal{A}$ is not.
However, this hidden edge mode $\tilde\varphi$ does not play any role in the dynamics because it is a gauge degree of freedom.
We have introduced it so that we have an explicit dual counterpart of the gauge charge \eqref{Q_e^Dalpha}.
The gauge transformation may well be understood as
\begin{align}
    \mathcal{A} \to\mathcal{A}+ d\tilde\alpha.
\end{align}
Nevertheless, $\mathcal{A}$ is not a dynamical variable and we thus should interpret $d\mathcal{A}$ as $d(\mathcal{A}+d\tilde\varphi)$ and consider the transformation to be the shift of $\tilde{\varphi}$ as \eqref{eq:hidden magnetic symmetry}.

One may consider that we can introduce the edge mode $\tilde\varphi$ even in the Dirichlet case $S_D$, by regarding $d\mathcal{A}$ as $d(\mathcal{A}+d\tilde\varphi)$.
It is possible, though one can see that the edge modes $\varphi$ and $\tilde\varphi$ appear only in the form of
\begin{align}
    \frac{1}{2\pi}\int_{\del\Sigma}d(\delta(\varphi-\tilde\varphi)\delta\hat{a})
\end{align}
which implies that the vector-like transformation
\begin{align}
\tilde\varphi\to\tilde\varphi+\alpha, \quad\varphi\to\varphi+\alpha
\end{align}
is degenerate in this sense.
In other words, these transformations are not independent on the physical phase space.
We thus do not need to introduce such an unnecessary gauge mode $\tilde{\varphi}$.

\subsection{Singular transformation as an insertion of line operator}\label{sec:singular}
The gauge parameters might be taken singular as considered in \cite{Hosseinzadeh:2018dkh, Freidel:2018fsk}.
It motivates us to consider singular gauge transformations. 
We will see that, for the singular transformation, charges \eqref{eq:charge Q_e^+} and \eqref{eq:charge Q_m^-} can be non-vanishing. 
Another possibility, in which the gauge field itself is taken to be singular, is discussed in \cref{subsec:QED?}.

Not every singular transformation is allowed because most transformations break the condition of no current at the boundary $d*F\overset{\Delta}{=}0$, which we impose in our analysis. We thus focus on a specific class of singularities described as follows.
Consider a two-dimensional plane with the polar coordinates $(r,\phi)$ and a 1-form $d\phi$.
This 1-form satisfies the closed condition $d(d\phi)=0$ away from the origin, while the integral around the origin $\oint d\phi$ gives $2\pi$.
Hence, one finds
\begin{align}
    d(d\phi) = 2\pi\delta^{(2)}(x,y)dx\wedge dy,
\end{align}
where $x,y$ are the Cartesian coordinates.
Generalizing it, we may consider a 1-form $d\alpha_C$ on three-dimensional space $\Delta$ satisfying
\begin{align}
    d(d\alpha_C) = 2\pi \delta_C
\end{align}
where $C$ is a curve in $\Delta$, and $\delta_C$ is a kind of the Poincar\'e dual, which gives
\begin{align}
    \int_\Delta\mathcal{A}\wedge\delta_C = \int_C \mathcal{A}.
\end{align}

This kind of transformation with a singular parameter $\alpha_C$ provides us with a non-vanishing, or physical, charge.
Let us see that a singular transformation $\alpha_C$ for a closed loop $C$ in $\Delta$ does not spoil the boundary condition $d*F\overset{\Delta}{=}0$.
To begin with, note that $\delta_C$ is closed when $C$ is a closed loop.
One can easily check it by taking a test 0-form $f$ and integrating
\begin{align}
    \int_\Delta f\cdot d\delta_C = -\int_\Delta df\wedge\delta_C=-\oint_C df = 0,
\end{align}
if $f$ is regular.
Thus, although $*F$ is shifted by $\frac{e^2}{2\pi}\delta_C$ under this transformation in the Neumann or Dirichlet boundary condition, $d*F\overset{\Delta}{=}0$ still holds as long as the curve $C$ is taken to be closed.
If $C$ is open, $d*F\overset{\Delta}{=}0$ is broken, which means that this transformation produces charges at the boundary.

Under a singular transformation of $\tilde\varphi$ in the Neumann boundary condition and that of $\varphi$ in the Dirichlet boundary condition with a closed loop $C$ in $\Delta$, each action is changed as
\begin{align}
    S_N\to S_N - \oint_C a,\quad S_D\to S_D - \oint_C \hat{a}.
\end{align}
One can immediately identify the first one with the Wilson loop of the dressed gauge field (see also \cite{Mathieu:2020fwg}), while, with the boundary condition \eqref{eq:boundary conditions with the S-wall}, one can see the second one as the Wilson loop of the dressed dual gauge field, or the 't Hooft loop.
In other words, singular transformations are nothing but insertions of the Wilson/'t Hooft loops on the boundary.
Although it is doubtful whether they should be regarded as symmetries of the theory since the actions are not invariant under these singular transformations, they may be useful for understanding boundary dynamics of gauge theories.
We leave it for future work.

We also note that if we perform a singular shift $\varphi \to \varphi +\alpha_C$ for the Neumann theory, the action $S_N$ is shifted by $\oint_C \mathcal{A}$. However, it is just a number without dynamical variables.
The associated Noether charge also vanishes on the physical phase space.  
It means that the singular transformation is also gauge redundancy. 
Similarly, the shift $\hat{\varphi}$ in $S_D$ is gauged even for singular parameters.

\subsection{What if QED?}\label{subsec:QED?}
Once charged matter is introduced to the system, electromagnetic duality and the electric $U(1)$ 1-form symmetry in the bulk are explicitly broken.
Nevertheless, the symplectic form is essentially unchanged as can be seen from Appendix~\ref{sec:symplectic form with charge}.
Then, the previous discussion about charges still holds even though the electromagnetic duality in the bulk is explicitly broken.
In the previous section, we considered singular transformations to obtain non-vanishing charges $Q_e^D[\alpha]$ for the Dirichlet boundary condition.  
In QED\footnote{In this paper, the Maxwell theory refers to a pure electromagnetic theory with the bulk action \eqref{action:Maxwell} and thus the bulk EoM is $d\ast F=0$. 
In QED, the gauge field can couple to a current which may be either dynamical or a fixed background, and thus $d\ast F$ can be non-vanishing.}, we can obtain non-vanishing $Q_e^D[\alpha]$ without introducing a singular parameter $\alpha$. 
Indeed, in QED, there are sectors where the total charge operator
\begin{align}
    Q = \frac{1}{e^2}\int_{\partial\Sigma}*F=\frac{1}{2\pi }\int_{\partial\Sigma} d\hat{a}
\end{align}
does not vanish. 
It means we have to require that $\hat{a}$ be a nontrivial connection.\footnote{In the Maxwell theory, $\hat{a}$ cannot be nontrivial because $Q=0$.}
Thus, one can obtain non-vanishing charges $Q_e^D[\alpha]$ without considering a singular parameter $\alpha$ for the Dirichlet boundary condition.

On the other hand, for the Neumann boundary condition, we have to consider a singular transformation for $\tilde\varphi$ in order to obtain non-vanishing $Q_m^N$ if there is no monopole. 
If we allow monopoles, $a$ can be a nontrivial connection and then $Q_m^N[\tilde\alpha]$ becomes non-vanishing even for regular $\tilde\alpha$.

One may suspect that, even if we have charged matter in the bulk, there is \textit{boundary electromagnetic duality}\footnote{
Here, we mean by the boundary electromagnetic duality the self-duality under an exchange of $F$ and $*F$ \textit{only} at the boundary without changing the bulk EoM.
One can, of course, obtain another solution if one allows bulk transformations and monopoles in the bulk, but it changes the bulk EoM $d*F=*j_e$ and $dF=0$ into $d*F=0$ and $dF=*j_m$.
} which exchanges $F$ and $*F$ at the boundary since $d*F\overset{\Delta}{=}0$ is satisfied as well as $dF=0$, as discussed in \cite{Freidel:2018fsk}.
However, we do not have such a duality.
One can confirm it at the classical level as follows.
First of all, solve the EoM $d*F=*j$ under the Neumann condition $*F|_\Delta=\frac{e^2}{2\pi}d\mathcal{A}$.
Then we obtain $A$ in the bulk, and the boundary gauge field $A|_{\Delta}$ and its derivative $F|_\Delta=dA|_\Delta$.
If the boundary duality really exists, exchanging $F|_{\Delta}$ and $*F|_{\Delta}$ must give another solution.
But we cannot generally solve the EoM with both these boundary conditions imposed because this problem is too constrained.
One might expect that edge modes resolve this issue; however, our argument is written solely in terms of $F$ and $*F$, which are insensitive to edge modes. 
The edge modes are just gauge degrees of freedom in the Neumann boundary condition, and they are not relevant to the dynamics. 
Consequently, exchanging $F$ and $*F$ at the boundary is not a genuine duality.
We can also confirm it from the viewpoint of the S-wall.
If the boundary electromagnetic duality exists, there is a topological operator across which the boundary condition is changed, for example, from the Dirichlet to the Neumann boundary condition. 
This would correspond to removing the $W$ part of the $S$-wall in Fig.~\ref{fig:bending wall operator}. Then, there is a boundary of the S-wall, and we cannot move it arbitrarily because the end of the topological defect is not topological in general. 
Thus, we do not have a topological operator implementing the boundary electromagnetic duality.

\subsection{Relation to previous work on boundary electromagnetic duality}
Before concluding this section, let us give a few words about existing literature.
Singular gauge transformations and edge modes are investigated in some literature.
In \cite{Hosseinzadeh:2018dkh}, singular gauge transformations are considered to define non-trivial charge in the Minkowski spacetime.
In \cite{Freidel:2018fsk} (and the following work \cite{Mathieu:2020fwg}), the authors studied them by observing the parameter $\hat\alpha$ ($\tilde\alpha$ in their notation) appears only in the form $d\hat\alpha$, not $\hat\alpha$ itself.
Their setup looks similar to ours, though they only considered a special case. 
The action considered in \cite{Freidel:2018fsk} is our Dirichlet action \eqref{eq:action under D} with $\mathcal{A}=0$ ($\hat a$ in this paper should be replaced by $\mathcal{A}$ in their notation), which leads to a boundary EoM $da \overset{\Delta}{=} 0$. Then, they use a symplectic form which does not follow straightforwardly from the action.\footnote{The \mbox{(pre-)symplectic} form obtained straightforwardly from the action \eqref{eq:action under D} is \eqref{symp-DwithEdge}, though the form used in \cite{Freidel:2018fsk} is \begin{align}
    \frac{1}{e^2}\int_\Sigma \delta A\wedge\delta*F +\frac{1}{2\pi}\int_{\del\Sigma}\delta a\wedge\delta\hat{a},
\end{align}
where this $\hat{a}$ is $\mathcal{A}$ in their notation. Then, it seems that there is a boundary duality exchanging $a$ and $\hat{a}$.
However, this should be understood just as a canonical transformation, not a genuine symmetry of the theory, which maps a state to another state quantum mechanically or a solution to another solution classically.
As argued above, we do not have a genuine boundary electromagnetic duality.} 
Also, the singular transformation investigated in \cite{Freidel:2018fsk, Mathieu:2020fwg} is associated with the magnetic edge mode $\hat\varphi$, or charge \eqref{eq:charge Q_m^+}, which is different from the transformations examined in this paper.
We also comment that, if we take  $\mathcal{A}$ to be a dynamical field in the Neumann theory \eqref{eq:action under N}, we have $F\overset{\Delta}{=} 0$ from the variation of $\mathcal{A}$, and thus it reduces to a specific Dirichlet boundary condition.

%%%%%%%%%%%%%%%
\section{Retake: modified boundary conditions}\label{sec:Retake}
None of the surface charges considered in the previous section are physical in the sense that they generate only gauge redundancies.
If we consider some singular transformations, we can obtain non-vanishing charges, though they are not symmetries because the action is not invariant.
This means that, under either Neumann or Dirichlet boundary conditions, there is no physical large ``gauge'' symmetry and no edge modes as firstly pointed out in \cite{Carrozza:2021gju} (see also \cite{Hoehn:2025pmx}).
It motivates us to consider other boundary conditions.

In this section, we modify boundary conditions as done in \cite{Ball:2024hqe, Araujo-Regado:2024dpr} to introduce physical edge modes.\footnote{The dynamical edge mode boundary condition in \cite{Ball:2024hqe} is extended to higher form gauge fields in \cite{Ball:2024xhf}, and it is related to \cite{David:2021wrw}. The dynamical edge mode boundary condition is also studied in \cite{Canfora:2024awy}.}
Then, we will find new boundary conditions dual to them.
We confirm that the associated physical charges can be regarded as topological operators.

\subsection{Modified boundary condition in Neumann}\label{subsec:Mod in N}
Let us first examine the Neumann theory with action \eqref{eq:action under N}.\footnote{We can instead consider \eqref{eq:action under N with mag edge} with the hidden edge mode $\tilde{\varphi}$. It is introduced to write down the magnetic gauge charge \eqref{eq:charge Q_m^-}, which is still a gauge charge even under modified boundary conditions considered in this paper.  
Thus, we omit $\tilde{\varphi}$ and focus on the action \eqref{eq:action under N}.}
We took the standard Neumann boundary condition $*F\overset{\Delta}{=}\frac{e^2}{2\pi}d\mathcal{A}$, and saw that under this condition the \mbox{(pre-)symplectic} form is degenerate along the large gauge transformations $A \to A +d \alpha$ or $\varphi \to \varphi + \alpha$, which means the large gauge transformation is also gauged.

To have a physical large gauge symmetry, it is necessary to impose another boundary condition.
Since the variation of the action \eqref{eq:action under N} is given by \eqref{deltaS_N_again} and is rewritten as
\begin{align}
    \delta S_N[A,\varphi;\mathcal{A}]
    = &-\frac{1}{e^2}\int_{\mathcal{M}}\delta A\wedge d*F +\frac{1}{e^2}\int_{\Delta}\delta a\wedge\ab(*F-\frac{e^2}{2\pi}d\mathcal{A}) + \frac{1}{e^2}\int_{\Delta}\delta\varphi\cdot d*F \notag\\
    & -\frac{1}{e^2}\int_{\mathcal{M}_{+\infty}-\mathcal{M}_{-\infty}}\delta A\wedge*F - \frac{1}{e^2}\int_{\Delta_{+\infty}-\Delta_{-\infty}}\delta\varphi*F,
\end{align}
it is required, besides imposing the bulk EoM $d*F=0$, that the variation at the boundary should be a total derivative \cite{Harlow:2019yfa}
\begin{align}
\label{bd-cond_dc}
    \frac{1}{e^2}\delta a\wedge\ab(*F-\frac{e^2}{2\pi}d\mathcal{A}) \overset{\Delta}{=} dc.
\end{align}
The left-hand side can be expressed in coordinates as
\begin{align}
    \frac{1}{e^2}\ab(\frac{1}{2}\delta a_t\ab(*F-\frac{e^2}{2\pi}d\mathcal{A})_{ab}+\delta a_{a}\ab(*F-\frac{e^2}{2\pi}d\mathcal{A})_{bt})\notag\\
    dt\wedge d\Omega^a\wedge d\Omega^b,
\end{align}
where $a=1,2$ denotes a label of angular variables.
Instead of the standard Neumann boundary condition, one may take
\begin{align}\label{eq:soft Neumann}
    \delta a_t \overset{\Delta}{=} 0,\quad \ab(*F-\frac{e^2}{2\pi}d\mathcal{A})_{bt} \overset{\Delta}{=} 0\quad (dc=0).
\end{align}
This is a kind of boundary condition referred to as dynamical edge mode boundary condition (DEM) \cite{Ball:2024hqe} or soft Neumann boundary condition \cite{Araujo-Regado:2024dpr}.

A \mbox{(pre-)symplectic} form under this boundary condition is also given by
\begin{align}
    \Omega_{N'}
    &= \frac{1}{e^2}\int_{\Sigma}\delta A\wedge\delta*F + \frac{1}{e^2}\int_{\del\Sigma}\delta\varphi\cdot\delta*F.
    \label{symp_modiN}
\end{align}
For the standard Neumann condition, the last term vanishes because $*F$ is fixed, while for the modified Neumann boundary condition $\delta*F$ does not vanish, since the $ab$-component of $*F$ is no longer constrained.
Specifically, we have a non-trivial symmetry transformation
\begin{align}
    \label{shift-vp-modiN}\varphi\to\varphi+\alpha,\quad \alpha=\alpha(\Omega),
\end{align}
with $A$ fixed.
Note that $\alpha$ must be independent of $t$ so that it is consistent with the boundary condition \eqref{eq:soft Neumann}.
The variation of the corresponding charge is computed from the form \eqref{symp_modiN} as
\begin{align}\label{delQ_e}
    \delta Q_e[\alpha]= \delta\ab(-\frac{1}{e^2}\int_{\partial\Sigma}\alpha*F).
\end{align}
Naively, one may conclude that the corresponding charge is given by
\begin{align}\label{NaiveQ}
    -\frac{1}{e^2}\int_{\partial\Sigma}\alpha*F.
\end{align}
However, this is not conserved in general since $d_\Delta (\alpha*F)\neq0$.
In order to make it conserved, we need to add an element of the kernel of $\delta$, and indeed, by noting $\delta(\frac{1}{2\pi}\alpha(\Omega)d\mathcal{A})=0$, we can find a conserved boundary current $J$ as\footnote{We have 
\begin{align}
    J^t &=-\frac{1}{2}\alpha(\Omega)\ab(\frac{1}{e^2}*F-\frac{1}{2\pi}d\mathcal{A})_{ab} E^{ab}, \label{Jt}\\
    J^a &=\alpha(\Omega)\ab(\frac{1}{e^2}*F-\frac{1}{2\pi}d\mathcal{A})_{tb} E^{ab},\label{Ja}
\end{align}
where $E^{ab}$ is a covariant volume form of $S^2$ ($E^{ab}=\frac{\epsilon^{ab}}{R^2 \sqrt{\gamma}}$).
For the boundary condition \eqref{eq:soft Neumann}, $J^a=0$.}
\begin{align}\label{currentJ}
    *_\Delta J[\alpha] = -\alpha(\Omega)\ab(\frac{1}{e^2}*F-\frac{1}{2\pi}d\mathcal{A}),
\end{align}
and a conserved charge
\begin{align}\label{eq:modified charge in modified N}
     Q_e[\alpha]= \int_{\del\Sigma}*_\Delta J[\alpha]= -\int_{\del\Sigma}\alpha(\Omega)\ab(\frac{1}{e^2}*F-\frac{1}{2\pi}d\mathcal{A}).
\end{align}
This charge is topological, and  we can arbitrarily modify the surface $\del\Sigma$ along the boundary $\Delta$, since $d_\Delta *_\Delta J[\alpha]=0$.\footnote{
To make naive one \eqref{NaiveQ} itself conserved, one has to require 
\begin{align}\label{eq:guarantee topness}
    \del_t(*F)_{ab}\varepsilon^{ab} = \del_t\ab(\sqrt{-g}F^{tr}) \overset{\Delta}{=}0,
\end{align}
which is imposed in soft boundary conditions \cite{Araujo-Regado:2024dpr}, or equivalently
\begin{align}
    \nabla_t\nabla_a\mathcal{A}_b\varepsilon^{ab}=0.
\end{align}
This condition is automatically satisfied if one chooses $d\mathcal{A}=0$ as in \cite{Ball:2024hqe}.
}
In addition, this charge \eqref{eq:modified charge in modified N} can take a non-zero value, while the electric gauge charge always vanishes in the standard Neumann boundary condition. 
It means that the shift symmetry \eqref{shift-vp-modiN} represents a physical symmetry.
Thus, the edge mode $\varphi$ represents a physical degree of freedom, as expected. 
This is because $*F$ is not completely frozen at the boundary under the boundary condition \eqref{eq:soft Neumann} in contrast to the standard Neumann one where $*F$ is fixed to $\frac{e^2}{2\pi}d\mathcal{A}$ at $\Delta$.

We have obtained a boundary topological operator \eqref{eq:modified charge in modified N} associated with the physical boundary symmetry \eqref{shift-vp-modiN}. It can be regarded as a kind of subsystem symmetry \cite{PhysRevB.66.054526, Chamon:2004lew, Haah:2011drr, Vijay:2015mka, Seiberg:2020bhn}.

One can also go to a picture without edge modes just by peeling off the dressing field $\varphi$, or setting $\varphi=0$ at each stage.
Then, we have the Maxwell theory with the modified Neumann boundary condition, and the theory has the physical large gauge symmetry
\begin{align}
    A\to A+d\alpha,\quad \alpha=\alpha(\Omega),
\end{align}
unlike the standard Neumann boundary condition where the large gauge transformation represents gauge redundancy.

One may take another boundary condition satisfying \eqref{bd-cond_dc}. We impose
\begin{align}
    \delta a_t \overset{\Delta}{=} 0,\quad \delta a_a \overset{\Delta}{=} \del_a\delta\lambda.
\end{align}
The remaining term in the requirement \eqref{bd-cond_dc} can be computed as 
\begin{align}
    \delta a_a\ab(*F-\frac{e^2}{2\pi}d\mathcal{A})_{bt}\varepsilon^{ab}
    &= \del_a\delta\lambda \ab(*F-\frac{e^2}{2\pi}d\mathcal{A})_{bt}\varepsilon^{ab} \sim - \delta\lambda\del_a\ab(*F-\frac{e^2}{2\pi}d\mathcal{A})_{bt}\varepsilon^{ab}  \notag\\
    &\propto -\delta\lambda\del_t\ab(*F-\frac{e^2}{2\pi}d\mathcal{A})_{ab}\varepsilon^{ab}
\end{align}
where we dropped a total derivative term with respect to angular variables since it does not contribute to variations, and we used the EoM in going to the last line from the previous line.
Thus, to make the variational principle well-defined, we further impose\footnote{
Of course, this is not necessarily required to be vanishing but a total derivative.
However, the large gauge symmetry in that case is found to be gauged again.
}
\begin{align}
    \del_t\ab(*F-\frac{e^2}{2\pi}d\mathcal{A})_{ab}\varepsilon^{ab}\overset{\Delta}{=}0.
\end{align}
After all, the alternative boundary condition is given by
\begin{align}\label{eq:soft Dirichlet}
    \delta a_t \overset{\Delta}{=} 0,\quad \delta a_a \overset{\Delta}{=} \del_a\delta\lambda,\quad \del_t\ab(*F-\frac{e^2}{2\pi}d\mathcal{A})_{ab}\varepsilon^{ab}\overset{\Delta}{=}0,
\end{align}
where $dc=0$.
This is the boundary condition referred to as the soft Dirichlet boundary condition \cite{Araujo-Regado:2024dpr}.
Note that the third condition in \eqref{eq:soft Dirichlet} is also satisfied in the boundary condition \eqref{eq:soft Neumann} automatically.

In this boundary condition \eqref{eq:soft Dirichlet},  the shift transformation
\begin{align}
    \varphi\to\varphi+\alpha,\quad \alpha=\alpha(\Omega)
\end{align}
is also a physical symmetry as in the case of \eqref{eq:soft Neumann}, giving the same conserved charge as \eqref{eq:modified charge in modified N}.
More precisely, the current $\eqref{currentJ}$ is not conserved in this boundary condition, and we have to add another kernel of $\delta$. 
The conserved boundary current is given by\footnote{If we use \eqref{currentJ}, we obtain \eqref{Jt} and \eqref{Ja}, and $J^a$ does not vanish under the boundary condition \eqref{eq:soft Dirichlet}. However, we have more freedom to add a kernel of $\delta$ in \eqref{delQ_e} using $\delta \int_{\partial\Sigma} \nabla_a(\cdots)=0$, since $S^2$ has no boundary.}
\begin{align}
    J^t=-\frac{1}{2}\alpha(\Omega)\ab(\frac{1}{e^2}*F-\frac{1}{2\pi}d\mathcal{A})_{ab}E^{ab}, \quad J^a=0,
\end{align}
where $E^{ab}$ is a covariant volume form of $S^2$ ($E^{ab}=\frac{\epsilon^{ab}}{R^2 \sqrt{\gamma}}$).
For this improved $J$, we have $d_\Delta *_\Delta J[\alpha]=0$. 
Thus, under the boundary condition \eqref{eq:soft Dirichlet}, we also have a boundary topological operator \eqref{eq:modified charge in modified N}.

Again, we can go back to the picture without edge modes by eliminating $\varphi$ at each stage and we obtain the physical large gauge symmetry
\begin{align}
    A\to A+d\alpha,\quad \alpha=\alpha(\Omega).
\end{align}

Finally, it should be noted that we are not aware of boundary conditions that make the magnetic charge \eqref{eq:charge Q_m^-} non-vanishing without introducing magnetic matter.
Therefore, this charge should be understood only as a formal structure dual to the Dirichlet case.
Once we obtain physical charges, or operators, $Q_e$ and $Q_m$, we can consider a commutator between them $[Q_e[\alpha],Q_m[\hat\alpha]]$. It is reported that this commutator has a central charge \cite{Freidel:2018fsk,Geiller:2021gdk}.
However, to make sense of the central charge, both charges have to be physical;
otherwise the central charge trivially vanishes on the physical space.
In the boundary conditions considered in our paper, we have not found a situation in which both charges are simultaneously physical. 
It is an interesting question whether boundary conditions exist such that we have a nontrivial central charge. Additional boundary degrees of freedom may provide a possible way to realize such a situation.
We leave this issue for future work.

\subsection{Modified boundary condition in Dirichlet}
We have seen that for the action \eqref{eq:action under N} we can take the modified boundary conditions \eqref{eq:soft Neumann} or \eqref{eq:soft Dirichlet} instead of the standard Neumann boundary condition.
Unlike the standard Neumann condition, the modified ones allow a physical edge mode and physical boundary 0-form symmetry associated with the shift of the edge mode.
We next consider the action \eqref{eq:action under D}, which can be regarded as adding the S-wall into \eqref{eq:action under N}.
It is expected that we have dual counterparts of the boundary conditions \eqref{eq:soft Neumann} or \eqref{eq:soft Dirichlet}.

The variation of \eqref{eq:action under D} is evaluated as
\begin{align}
    \delta S_D[A,\hat{A},\varphi,\hat\varphi;\mathcal{A}]
    &=-\frac{1}{e^2}\int_{\mathcal{M}}\delta A\wedge d*F + \frac{1}{e^2}\int_{\Delta}\delta a\wedge\ab(*F-\frac{e^2}{2\pi}d\hat{a})\notag \\
    &\quad -\frac{1}{2\pi}\int_{\Delta}(da-d\mathcal{A})\wedge\delta\hat{a} + \frac{1}{e^2}\int_{\Delta}\delta\varphi\cdot d*F \notag\\
    &\quad - \frac{1}{e^2}\int_{\mathcal{M}_{+\infty}-\mathcal{M}_{-\infty}}\delta A\wedge*F - \int_{\Delta_{+\infty}-\Delta_{-\infty}}\ab(\frac{1}{e^2}\delta\varphi*F -\frac{1}{2\pi}a\wedge \delta\hat{a}).
\end{align}
In order to make the theory well-defined, it is required that
\begin{align}
    \frac{1}{e^2}\delta a\wedge\ab(*F-\frac{e^2}{2\pi}d\hat{a}) &\overset{\Delta}{=} dc,\label{variation_a_D}\\
    -\frac{1}{2\pi}(da-d\mathcal{A})\wedge\delta\hat{a} &\overset{\Delta}{=}dc',\label{variation_hat-a_D}
\end{align}
in addition to the bulk EoM.
From the viewpoint of the S-wall, the boundary gauge field $\hat{a}$ in $S_D$ corresponds to $a$ in $S_N$, and thus we impose, on $\hat{a}$, a boundary condition which is dual to the modified boundary condition \eqref{eq:soft Neumann} or \eqref{eq:soft Dirichlet}.
From \eqref{variation_hat-a_D}, we require
\begin{align}\label{eq:dual soft Neumann temp}
    \delta\hat{a}_t\overset{\Delta}{=}0,\quad (da-d\mathcal{A})_{bt}\overset{\Delta}{=}0\quad (dc'=0)
\end{align}
corresponding to \eqref{eq:soft Neumann}, or
\begin{align}\label{eq:dual soft Dirichlet temp}
    \delta\hat{a}_t\overset{\Delta}{=}0,\quad \delta \hat{a}_a\overset{\Delta}{=}\del_a\delta \lambda, \quad \del_t(da-d\mathcal{A})_{ab}\varepsilon^{ab}\overset{\Delta}{=}0,
\end{align}
with $dc'=0$, corresponding to \eqref{eq:soft Dirichlet}.

Let us discuss the boundary condition \eqref{eq:dual soft Neumann temp}.
A similar analysis to \eqref{eq:dual soft Dirichlet temp} can be carried out as well and yields the same expression of charges.
\eqref{eq:dual soft Neumann temp} leads to a constraint on the variation of $a$ as
\begin{align}\label{eq:decomposition of delta a}
    \delta a \overset{\Delta}{=}\delta \Lambda_a d\Omega^a + d\delta\lambda,\quad \delta\Lambda_a=\delta\Lambda_a(\Omega).
\end{align}
Substituting it for \eqref{variation_a_D}, we demand either
\begin{align}\label{bd-cond-Lambda-or-F}
    \delta \Lambda_a=0,\ \text{or}\ \ab(*F-\frac{e^2}{2\pi}d\hat{a})_{bt}\overset{\Delta}{=}0.
\end{align}
In both cases, we have
\begin{align}
    \frac{1}{e^2}\delta a\wedge\ab(*F-\frac{e^2}{2\pi}d\hat{a}) &\overset{\Delta}{=} d\ab(\frac{1}{e^2}\delta\lambda\ab(*F-\frac{e^2}{2\pi}d\hat a))
\end{align}
and obtain a modified \mbox{(pre-)symplectic} form
\begin{align}
    \Omega_{D'}
    &= \frac{1}{e^2}\int_\Sigma \delta A\wedge\delta*F + \frac{1}{e^2}\int_{\del\Sigma}\delta\varphi\cdot\delta*F \notag \\
    &\quad +\frac{1}{2\pi}\int_{\del\Sigma}\delta a\wedge\delta\hat{a} - \frac{1}{e^2}\int_{\del\Sigma}\delta\lambda\cdot\delta\ab(*F-\frac{e^2}{2\pi}d\hat a).
\end{align}
It can be rewritten by means of \eqref{eq:decomposition of delta a} as
\begin{align}
    \Omega_{D'}
    &= \frac{1}{e^2}\int_\Sigma \delta A\wedge\delta*F + \frac{1}{e^2}\int_{\del\Sigma}\delta\varphi\cdot\delta*F \notag \\
    &\quad +\frac{1}{2\pi}\int_{\del\Sigma}\delta \Lambda\wedge\delta\hat{a} - \frac{1}{e^2}\int_{\del\Sigma}\delta\lambda\cdot\delta*F.
\end{align}
Under the electric transformation $\varphi\to\varphi+\alpha$, only $\lambda$ is shifted by $\alpha$, and it does not change the \mbox{(pre-)symplectic} form.
That is, this transformation remains gauged.
On the other hand, there is a nontrivial magnetic transformation
\begin{align}
    \label{magnetic-shift}
    \hat\varphi\to\hat\varphi+\hat\alpha,\quad \hat\alpha = \hat\alpha(\Omega).
\end{align}
If we take the latter condition in \eqref{bd-cond-Lambda-or-F}, we obtain the variation of the magnetic charge as
\begin{align}
    \delta Q_m[\hat\alpha] = \delta\ab(\frac{1}{2\pi}\int_{\del\Sigma}a\wedge d\hat\alpha) = \delta\ab(\frac{1}{2\pi}\int_{\del\Sigma}\hat\alpha F).
\end{align}
In order to obtain the conserved current and charge, we need to use the kernel of $\delta$ and we can indeed construct a conserved boundary magnetic current as
\begin{align}\label{astJm}
    *_\Delta J_m[\hat\alpha] = \frac{1}{2\pi}\hat\alpha(\Omega) (F -d\mathcal{A})
\end{align}
and a conserved magnetic charge
\begin{align}\label{Qm-Diri}
    Q_m[\hat\alpha] = \frac{1}{2\pi}\int_{\del\Sigma} \hat\alpha(\Omega)(F-d\mathcal{A}).
\end{align}
Note that, if we take the former condition in \eqref{bd-cond-Lambda-or-F}, the magnetic symmetry \eqref{magnetic-shift} is also gauged.

The topologicality of the charge $Q_m[\hat\alpha]$ is guaranteed by $d_\Delta*_\Delta J_m[\hat\alpha]=0$.
While the magnetic charge disappears in the standard Dirichlet boundary condition, this charge is non-vanishing, implying that the magnetic edge mode $\hat\varphi$ is a physical degree of freedom and its shift symmetry is a physical symmetry.
The reason for this change is that $F$ has an unconstrained component at the boundary under boundary conditions \eqref{eq:dual soft Neumann temp} and \eqref{eq:dual soft Dirichlet temp} unlike the standard Dirichlet boundary condition where $F$ is fixed to $d\mathcal{A}$.
Note that if we use the boundary condition \eqref{eq:dual soft Dirichlet temp} instead of \eqref{eq:dual soft Neumann temp},
the current $J_m$ is not \eqref{astJm}. We need further improvement as in the Neumann case with the boundary condition \eqref{eq:soft Dirichlet}. After the improvement, the conserved current is given by $J_m^t=\frac{1}{4\pi}\hat\alpha(\Omega) (F -d\mathcal{A})_{ab}E^{ab}$ and $J_m^a=0$.

The charge $Q_m[\hat{\alpha}]$ and the boundary conditions we have considered are indeed dual to the charge $Q_e[\alpha]$ and the boundary conditions in the Neumann case.
Recall the action with the S-wall requires $*F\overset{\Delta}{=}\frac{e^2}{2\pi}d\hat{a}$ as in \eqref{eq:boundary conditions with the S-wall}.
This implies that one should identify the $\hat{a}$ with a dual gauge field at least on the boundary.
This identification and the fact that the S-wall exchanges the gauge field and the dual gauge field, up to gauge, help us understand the duality: the S-wall maps the modified boundary conditions \eqref{eq:soft Neumann} and \eqref{eq:soft Dirichlet} to \eqref{eq:dual soft Neumann temp} and \eqref{eq:dual soft Dirichlet temp} respectively.
At the same time, the electric charge $Q_e[\alpha]$ and electric edge mode $\varphi$ are mapped to the magnetic charge $Q_m[\hat\alpha]$ and magnetic edge mode $\hat\varphi$ by the S-wall.
As before, we can throw away edge modes by eliminating $\varphi$ and $\hat\varphi$ at each stage to have a large gauge symmetry
\begin{align}
    \hat A\to \hat A+d\hat\alpha.
\end{align}

Note that we have not been able to find a non-vanishing electric charge \eqref{eq:charge Q_e^+} in the Dirichlet action $S_D$ as well as the magnetic charge in the Neumann action $S_N$.
It is necessary to find a boundary condition which makes it non-vanishing in order to justify the existence of a central charge studied in \cite{Freidel:2018fsk, Geiller:2021gdk} since the setup used in these papers can be obtained by setting $d\mathcal{A}=0$ in our $S_D$.

\section{Summary and discussion}\label{sec:summary}

In this paper, we studied boundary symmetries of four-dimensional Maxwell theory and QED with a finite boundary from the viewpoint of topological operators and electromagnetic duality, motivated by the dual structure of the asymptotic charges in the case of asymptotic boundaries.
We reviewed the S-wall implementing the electromagnetic duality, and discussed the relation between the Neumann and Dirichlet boundary conditions, and also showed that inserting the S-wall along the boundary maps the theory with the standard Neumann boundary condition to the theory with the standard Dirichlet boundary condition.
We then studied the large gauge transformations, or shift transformations of edge modes for the Neumann and Dirichlet boundary conditions.
We can formally write down the electric surface charges for the Neumann case, and the electric and magnetic ones for the Dirichlet case, although all of them generate degenerate directions of the \mbox{pre-symplectic} forms.
Furthermore, the magnetic surface charge looks absent in the Neumann case, which is dual to the electric ones in the Dirichlet case. 
We succeeded in making it manifest by introducing the hidden magnetic edge mode.
Since the electric surface charge in the Dirichlet case and the magnetic one in the Neumann condition are total derivatives, we were motivated to consider singular transformations as in \cite{Hosseinzadeh:2018dkh, Freidel:2018fsk}, which turned out to be equivalent to insertions of the Wilson and 't Hooft loops on the boundary as suggested in \cite{Mathieu:2020fwg}.

Since all charges for large gauge transformations, either electric or magnetic, generate degenerate directions of the \mbox{pre-symplectic} forms under the standard Neumann or Dirichlet boundary conditions, the large gauge transformations (or the corresponding shifts of edge modes) are not physical symmetries for these boundary conditions.
A lesson of this analysis is that large gauge transformations should not automatically be identified with physical boundary symmetries. 
Whether a large gauge transformation is physical or a gauge redundancy is determined by the boundary condition and by the resulting \mbox{pre-symplectic} form.

This point also clarifies the role of edge modes. Introducing edge modes makes gauge invariance manifest by dressing the boundary gauge fields and is useful for discussing symmetries without separating small and large gauge transformations. However, edge modes are not necessarily physical degrees of freedom. Under the standard Neumann and Dirichlet boundary conditions, their shifts remain degenerate directions of the \mbox{pre-symplectic} form, and thus, the edge modes are just gauge degrees of freedom.

It motivated us to re-examine the boundary conditions.
We could consider not only the modified boundary conditions found in \cite{Ball:2024hqe, Araujo-Regado:2024dpr} associated with the Neumann action but also the dual one associated with the Dirichlet action.
In the Neumann case, the electric surface charge generates a physical boundary symmetry, and the charge is a non-trivial topological operator.
In the Dirichlet case, the magnetic surface charge is physical, and it is a non-trivial topological operator.

An open problem is to find boundary conditions under which both electric and magnetic surface charges are simultaneously physical. This is necessary for giving a physical interpretation to the central charge appearing in the algebra of electric and magnetic boundary charges discussed in \cite{Freidel:2018fsk, Geiller:2021gdk}.

The charges obtained in this work also suggest a broader viewpoint: one may regard boundary symmetries as generalized symmetries localized on the boundary. 
The boundary symmetry obtained in \cref{sec:Retake} is a 0-form symmetry at the boundary. 
The charges are supported on codimension-one surfaces, and thus they are codimension-two surface operators from the bulk perspective.
More generally, it would be natural to consider boundary $p$-form symmetries, whose charges are topological operators supported on codimension-$(p+1)$ surfaces of the boundary, or equivalently on codimension-$(p+2)$ surfaces from the bulk perspective.
As we saw, these boundary generalized symmetries may depend sensitively on the choice of boundary condition. 
It would be interesting to develop a systematic classification of boundary conditions in terms of the boundary generalized symmetries they support. 
A natural asymptotic counterpart is to reinterpret asymptotic symmetries and soft charges as generalized symmetries living on the asymptotic boundary. This may clarify the distinction between physical asymptotic symmetries, gauge redundancies, and singular transformations corresponding to line-operator insertions.
It is also interesting to consider generalized asymptotic symmetry (see, e.g., \cite{Afshar:2018apx} where asymptotic symmetries of higher-form gauge fields are investigated).

It is interesting to ask what happens if we add the theta term $S_{\theta}=\frac{\theta}{8\pi^2}\int_{\mathcal{M}}F\wedge F$.
The variational principle requires $\frac{1}{e^2}\delta a\wedge(*F - \frac{e^2\theta}{4\pi^2}F-\frac{e^2}{2\pi}d\mathcal{A})\overset{\Delta}{=}dc$. For the standard boundary condition $*F - \frac{e^2\theta}{4\pi^2}F-\frac{e^2}{2\pi}d\mathcal{A}=0$, the situation is similar to the case without the theta term, and we have no physical edge modes.
This remains true even when we consider the $SL(2,\mathbb{Z})$ dual of this boundary condition.
One can instead impose a modified boundary condition $\delta a_t=0,~(*F - \frac{e^2\theta}{4\pi^2}F-\frac{e^2}{2\pi}d \mathcal{A})_{bt}=0$, and we can also consider $SL(2,\mathbb{Z})$ dual of this boundary condition, which leads to another type of edge mode.
We will investigate these boundary conditions and edge modes in detail elsewhere.

\begin{figure}
    \centering
    \includegraphics[width=0.3\linewidth]{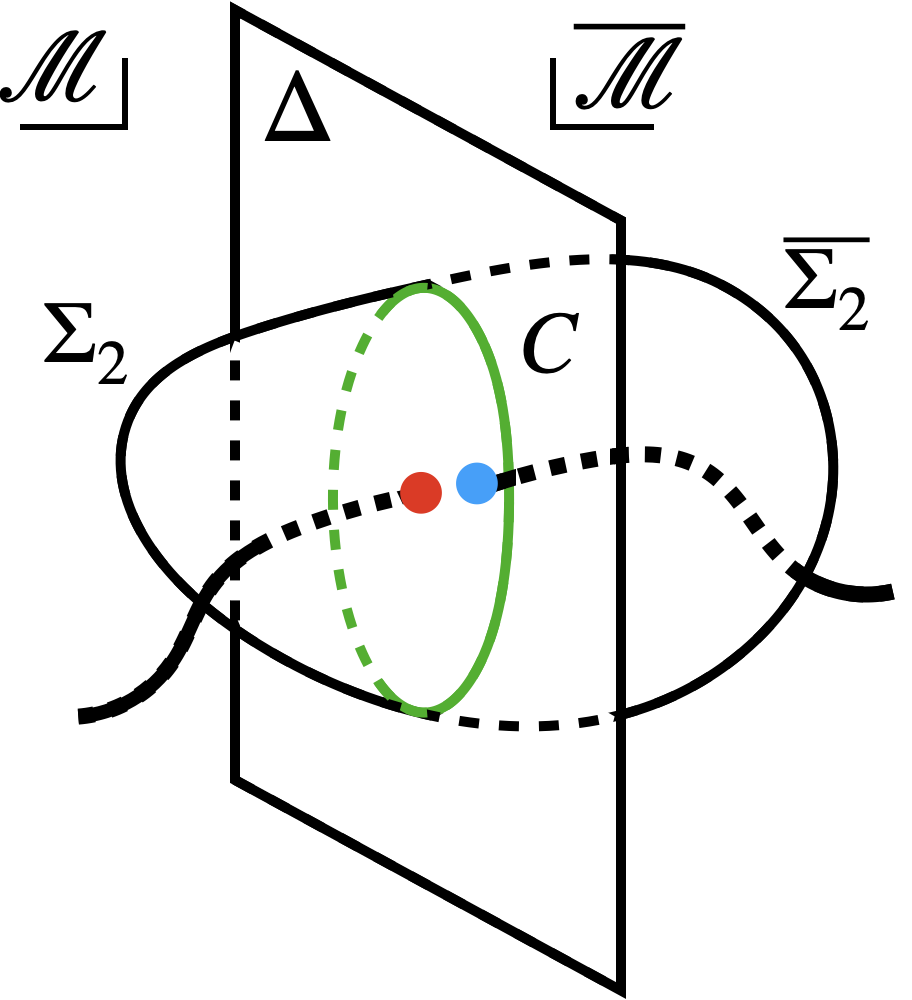}
    \caption{
    Suppose the ``outside'' of the spacetime referred to as $\overline{\mathcal{M}}$ separated from $\mathcal{M}$ by $\Delta$.
    A loop operator in the whole spacetime is cut by the boundary and an edge mode lives at the end.
    Similarly, a $U(1)$ 1-form operator defined on $\Sigma_2\cup_{C(\subset\Delta)}\overline{\Sigma_2}$ is also cut by the boundary to give a loop $C$ on the boundary.
    }
    \label{fig:origin of symmetry}
\end{figure}
Edge modes are crucial in various areas of physics as mentioned in the introduction, and they can often be associated with endpoints of Wilson loops on the boundary, which cannot exist in the bulk theory to ensure gauge invariance.
From this perspective, we can see the magnetic symmetry or edge mode $\hat\varphi$ as a remnant of the 't Hooft lines attached to the boundary, although there have not been any studies of edge modes associated with 't Hooft lines as far as we know.
It may be possible to regard the hidden edge mode we found in this paper as the remnant as well.
In addition, inspired by this observation, it is interesting to pursue other realizations of objects or degrees of freedom on the boundary arising from extended objects cut by boundaries in gauge theories other than the Wilson loops.
For example, the $U(1)$ 1-form symmetry operators are two-dimensional objects in Maxwell theory and QED, and they are reduced to the Wilson and 't Hooft loops on the boundary (see \cref{fig:origin of symmetry}).
This line of research may shed new light on gauge theories with boundaries.

%%%%%%%%%%%%%%%%%%
%%%%%%%%%%%%%%%%%%
%%%%%%%%%%%%%%%%%%
\acknowledgments
We thank Yuta Hamada, Philipp A. H\"ohn, Puttarak Jai-akson, Goncalo Araujo Regado, and Bilyana Tomova for valuable discussions. This research benefited from discussions at the workshop “Concepts of Quantum and Spacetime (KEK, Japan, March 2026)”.
KS is also grateful to Tsubasa Oishi for enlightening him about generalized symmetry.
KS was supported by JST SPRING, Grant Number JPMJSP2110, as well as Fujitsu Limited through Division of Graduate Studies Donor Designated Scholarship.
SS acknowledges support from JSPS KAKENHI Grant Numbers JP21K13927, JP22H05115, 	26K17145 and JST BOOST Program Japan Grant Number JPMJBY24E0.

\appendix
\begin{center}
\textbf{Appendix}
\end{center}
\section{Symplectic form and charge}\label{sec:symplectic form and charge}
In this appendix, we first derive various charges associated with symmetry transformations of the Maxwell actions with the Neumann and Dirichlet boundary conditions.
Then, we comment on the case of QED although expressions of boundary charges are not changed.

\subsection{Without charged matter}
We first consider the following two actions without edge modes: 
\begin{align}\label{eq:Neumann action}
    S_N[A;\mathcal{A}]= -\frac{1}{2e^2}\int_{\mathcal{M}}F\wedge*F - \frac{1}{2\pi}\int_{\Delta} A\wedge d\mathcal{A},
\end{align}
and
\begin{align}\label{eq:Dirichlet action}
    &S_D[A,\hat{A};\mathcal{A}]= -\frac{1}{2e^2}\int_{\mathcal{M}}F\wedge*F - \frac{1}{2\pi}\int_{\Delta} (A\wedge d\hat{A}-\hat{A}\wedge d\mathcal{A}),
\end{align}
where $N$ and $D$ denote the Neumann and Dirichlet, respectively. 
In the actions, $\mathcal{A}$ is a background and not the dynamical field.
These actions correspond to $S_-$ and $S_+$ in \cref{sec:elemag dual} respectively.

The variation of \eqref{eq:Neumann action} is given by
\begin{align}
    &\delta S_N[A;\mathcal{A}]
    \notag \\
    &= -\frac{1}{e^2}\int_{\mathcal{M}}\delta A\wedge d*F + \frac{1}{e^2}\int_{\Delta}\delta A\wedge\ab(*F-\frac{e^2}{2\pi}d\mathcal{A}) -\frac{1}{e^2}\int_{\mathcal{M}_{+\infty}-\mathcal{M}_{-\infty}}\delta A\wedge*F,
\end{align}
where $\mathcal{M}_{\pm\infty}$ and $\Delta_{\pm\infty}$ denote time-like infinities of each region, which implies a bulk EoM
\begin{align}
    \frac{1}{e^2}d*F~\hat{=}~0,
\end{align}
and a boundary condition\footnote{As discussed in \cref{{sec:Retake}}, we can take another boundary condition. In this appendix, we focus on the standard Neumann or Dirichlet boundary condition.}
\begin{align}
    \frac{1}{e^2}*F\overset{\Delta}{=}\frac{1}{2\pi}d\mathcal{A}.
\end{align}

Similarly, the variation of \eqref{eq:Dirichlet action} is
\begin{align}
    &\delta S_D[A,\hat{A};\mathcal{A}]\notag\\
    &=-\frac{1}{e^2}\int_{\mathcal{M}}\delta A\wedge d*F +\frac{1}{e^2}\int_{\Delta}\delta A\wedge\ab(*F-\frac{e^2}{2\pi}d\hat{a})
    -\frac{1}{2\pi}\int_{\Delta}(dA-d\mathcal{A})\wedge\delta\hat{A}\notag\\
    &\quad -\frac{1}{e^2}\int_{\mathcal{M}_{+\infty}-\mathcal{M}_{-\infty}}\delta A\wedge*F+\frac{1}{2\pi}\int_{\Delta_{+\infty}-\Delta_{-\infty}}A\wedge \delta\hat{a},
\end{align}
which implies the same bulk EoM
\begin{align}
    \frac{1}{e^2}d*F~\hat{=}~0,
\end{align}
and boundary conditions or boundary EoMs
\begin{align}
    \frac{1}{e^2}*F\overset{\Delta}{=}\frac{1}{2\pi}d\hat A,\quad F=dA\overset{\Delta}{=}d\mathcal{A}.
\end{align}

The \mbox{(pre-)symplectic} potentials for these actions can be read off as
\begin{align}
    \Theta_{N,\Sigma} &= -\frac{1}{e^2}\int_\Sigma \delta A\wedge*F,
\end{align}
and
\begin{align}
    \Theta_{D,\Sigma}
    &= -\frac{1}{e^2}\int_\Sigma \delta A\wedge*F + \frac{1}{2\pi}\int_{\del\Sigma}A\wedge\delta\hat A. 
\end{align}
Here, it should be noted that there are ambiguities in the definition of a \mbox{(pre-)symplectic} potential $\Theta$ as described in the main manuscript.
According to the construction developed in \cite{Harlow:2019yfa}, the boundary term of the variation of the actions does not need to vanish at the spatial boundary, but simply needs to be a total derivative $dc$, which modifies a symplectic potential and thereby a symplectic form.
Also, $c$ is determined up to a total derivative, which modifies a symplectic potential by a corner term.
The term is crucial when we define charges as in \cref{sec:elemag dual}.
%We performed partial integrals in the above variations using the ambiguity of $c$ and $dY$.

From the above \mbox{(pre-)symplectic} potentials, we obtain \mbox{(pre-)symplectic} forms
\begin{align}
    \Omega_{N,\Sigma} = \frac{1}{e^2}\int_\Sigma \delta A\wedge\delta*F
\end{align}
for the Neumann condition, and
\begin{align}
    \Omega_{D,\Sigma}= \frac{1}{e^2}\int_\Sigma \delta A\wedge\delta*F + \frac{1}{2\pi}\int_{\del\Sigma} \delta A\wedge\delta\hat A
\end{align}
for the Dirichlet condition.
The \mbox{(pre-)symplectic} form $\Omega_N$ looks gauge-variant for the large gauge transformations, but it is actually gauge invariant since $\delta*F\overset{\Delta}{=}0$. It implies that large gauge transformations are also gauge redundancy for these boundary conditions.

Now, we compute charges for the large transformations.
The charge corresponding to $A\to A+d\alpha$ is given by
\begin{align}\label{delQeN}
    \delta Q_e^N[\alpha] = -I_\alpha\Omega_N = \delta\ab(-\frac{1}{e^2}\int_{\del\Sigma}\alpha *F)
\end{align}
and
\begin{align}
    \delta Q_e^D[\alpha] = -I_\alpha\Omega_D = \delta\ab(-\frac{1}{2\pi}\int_{\del\Sigma}d(\alpha\hat A)).
\end{align}
Thus, we naively obtain 
\begin{align}\label{naiveQeN}
    Q_e^N[\alpha]=-\frac{1}{e^2}\int_{\del\Sigma}\alpha *F.
\end{align}
However, since $*F$ is fixed at the boundary as $*F\overset{\Delta}{=}\frac{e^2}{2\pi}d\mathcal{A}$ in the Neumann boundary condition, 
we have
\begin{align}
    \delta\ab(-\frac{1}{e^2}\int_{\del\Sigma}\alpha *F)=0.
\end{align}
This means that the large gauge transformations $A\to A+d\alpha$ are degenerate directions in the \mbox{(pre-)symplectic} form, and the transformations are gauge redundancy. 
$\alpha *F$ in \eqref{naiveQeN} is not dynamical, and is fixed by the background field $\mathcal{A}$. 
In this sense $Q_e^N[\alpha]$ is not an operator on the phase space. 
On the physical phase space, the gauge charge generating the gauge transformation vanishes, and indeed we obtain the vanishing gauge charge from the standard procedure constructing the Noether current.
We indeed have the freedom to add an element of the kernel of $\delta$ to $Q_e^N$ in \eqref{delQeN}, and by noting $\delta(\frac{1}{2\pi}\alpha d\mathcal{A})=0$, we can take 
\begin{align}
    Q_e^N[\alpha]=-\frac{1}{e^2}\int_{\del\Sigma}\alpha\left(*F-\frac{e^2}{2\pi}d\mathcal{A}\right)=0,
\end{align}
which vanishes from the boundary condition.
Similarly, we naively obtain 
\begin{align}\label{naiveQeD}
    Q_e^D[\alpha]=-\frac{1}{2\pi}\int_{\del\Sigma}d(\alpha\hat A).
\end{align}
It also vanishes for non-singular $\alpha$.

The charge corresponding to the large gauge transformation of $\hat A\to\hat A+d\hat\alpha$ for the action $S_D$ is given by
\begin{align}
    \delta Q_m^D[\hat\alpha] = -I_{\hat{\alpha}}\Omega_D = \delta\ab(\frac{1}{2\pi}\int_{\del\Sigma}A\wedge d\hat \alpha).
\end{align}
We then obtain naively
\begin{align}
   Q_m^D[\hat\alpha] = \delta\ab(\frac{1}{2\pi}\int_{\del\Sigma}A\wedge d\hat \alpha).
\end{align}
It leads to the expression \eqref{eq:charge Q_m^+}.
However, as explained in the main manuscript, by a partial integration, we have
\begin{align}
    \frac{1}{2\pi}\int_{\del\Sigma}A\wedge d\hat \alpha=\frac{1}{2\pi}\int_{\del\Sigma}\hat \alpha F,
\end{align}
and $F$ is fixed at the boundary $F\overset{\Delta}{=}d\mathcal{A}$ in the Dirichlet boundary condition. 
Thus, the large gauge transformation of $\hat A\to\hat A+d\hat\alpha$ is also degenerated in the \mbox{(pre-)symplectic} form as
\begin{align}
    -I_{\hat{\alpha}}\Omega_D = \delta\ab(\frac{1}{2\pi}\int_{\del\Sigma}A\wedge d\hat \alpha)=0.
\end{align}
Thus, similar to $Q_e^N[\alpha]$ for the Neumann boundary condition, $ Q_m^D[\hat\alpha]$ is just a number, and the gauge charge for $\hat A\to\hat A+d\hat\alpha$ vanishes on the physical phase space.

In the above explanation, we have not introduced the edge modes. Let us repeat the same argument with introducing edge modes as \eqref{eq:action under N} and \eqref{eq:action under D}.
The actions are 
\begin{align}
    S_N[A,\varphi;\mathcal{A}]
    = -\frac{1}{2e^2}\int_{\mathcal{M}}F\wedge*F - \frac{1}{2\pi}\int_\Delta a\wedge d\mathcal{A},
\end{align}
and
\begin{align}
    S_D[A,\hat{A},\varphi,\hat\varphi;\mathcal{A}]= -\frac{1}{2e^2}\int_{\mathcal{M}}F\wedge*F - \frac{1}{2\pi}\int_{\Delta} (a\wedge d\hat{a}-\hat{a}\wedge d\mathcal{A}),
\end{align}
which give the \mbox{(pre-)symplectic} forms under the Neumann and Dirichlet conditions
\begin{align}
    \Omega_{N}
    &= \frac{1}{e^2}\int_{\Sigma}\delta A\wedge\delta*F + \frac{1}{e^2}\int_{\del\Sigma}\delta\varphi\cdot\delta*F,
\end{align}
and
\begin{align}\label{symp-DwithEdge}
    \Omega_{D}
    = \frac{1}{e^2}\int_\Sigma \delta A\wedge\delta*F + \frac{1}{e^2}\int_{\del\Sigma}\delta\varphi\cdot\delta*F+\frac{1}{2\pi}\int_{\del\Sigma}\delta a\wedge\delta\hat{a}.
\end{align}

The Neumann theory has the gauge symmetry
\begin{align}
    A\to A+d\alpha,\quad \varphi\to\varphi-\alpha
\end{align}
with a vanishing charge
\begin{align}
    \delta Q_{\mathrm{gauge}} = -I_\alpha\Omega_N = 0.
\end{align}
It is also invariant under the shift transformation
\begin{align}
    \varphi\to\varphi+\alpha
\end{align}
with the corresponding charge
\begin{align}
    \delta Q_e = -I_\alpha\Omega_N = \delta\ab(-\frac{1}{e^2}\int_{\Sigma_2}\alpha*F).
\end{align}
However, it vanishes as in the case without edge modes, and thus the shift transformation is also the gauge redundancy.

The Dirichlet theory has the gauge symmetry
\begin{align}
    A\to A+d\alpha,\quad \varphi\to\varphi-\alpha
\end{align}
with a vanishing charge
\begin{align}
    \delta Q_{\mathrm{gauge}} = -I_\alpha\Omega_D = 0.
\end{align}
The theory is also invariant under the shift transformations
\begin{align}
    \varphi\to \varphi + \alpha,\quad \hat\varphi\to\hat\varphi+\hat\alpha
\end{align}
with the corresponding charges
\begin{align}
    \delta Q_e^D[\alpha] &= -I_\alpha\Omega_D = \delta\ab(-\frac{1}{2\pi}\int_{\del\Sigma}d(\alpha\hat a)), \\
    \delta Q_m^D[\hat\alpha] &= -I_{\hat{\alpha}}\Omega_D = \delta\ab(\frac{1}{2\pi}\int_{\del\Sigma}a\wedge d\hat \alpha).
\end{align}
They also vanish as in the case without edge modes. 
Thus, the shift transformations are also gauge redundancy, and the edge modes are gauge degrees of freedom.

\subsection{With charged matter}\label{sec:symplectic form with charge}
Let us add charged matter.
Adding a charged matter 
\begin{align}
    S_{N,D}\to S_{N,D} + S_{\mathrm{charge}}[\Phi] + \int_{\mathcal{M}}A\wedge*j
\end{align}
affects variations as
\begin{align}
    \delta S_{N,D}\to\delta S_{N,D} + \delta S_{\mathrm{charge}}[\Phi] + \delta\ab(\int_{\mathcal{M}}A\wedge*j).
\end{align}
By using the usual decomposition \cite{Iyer:1994ys}
\begin{align}
    \delta S_{\mathrm{charge}} = \int_{\mathcal{M}}(\delta\Phi\cdot E'[\Phi] + d\theta[\Phi,\delta\Phi]),
\end{align}
where $E'[\Phi]$ is an EoM in the absence of the gauge field, we obtain additional bulk EoMs
\begin{align}
    \frac{1}{e^2}d*F~\hat{=}~*j,\quad E[\Phi,A]~\hat=~0,
\end{align}
where $E[\Phi,A]$ is a matter EoM in the presence of the gauge field, and boundary conditions
\begin{align}
    \theta(\Phi,\delta\Phi)\overset{\Delta}{=}0,
\end{align}
where we ignore the boundary ambiguity of the symplectic potential for charged matter.
We impose that the current vanishes at the boundary $*j\overset{\Delta}{=}0$ for simplicity.
One simple choice of the boundary condition of the matter field $\Phi$ is the Dirichlet condition $\Phi\overset{\Delta}{=}0$.
For instance, the current is given by $j^\mu = \bar{\psi}\gamma^\mu\psi$ in QED and it vanishes at boundary when $\psi\overset{\Delta}{=}0$.

The expressions of the boundary charges are the same as the previous ones without charged matter under the assumption that $*j\overset{\Delta}{=}0$.
Indeed, the contribution of charged matter to the symplectic form is just an addition to that without them as
\begin{align}
    \Omega_{N,D}\to\Omega_{N,D} + \Omega_{\mathrm{charge}},
\end{align}
which gives the same expression of charges as the previous ones, although their analytic properties are different as discussed in the main part.
Finally, it should be noted that in this section we implicitly assume that the current does not contain derivatives of matter fields such as spinor QED (not scalar QED).

%\clearpage
%%%%%%%%%%%%%%%%%%%%%
%\bibliographystyle{apsrev4-2}
\bibliography{ref}
\bibliographystyle{JHEP.bst}
%%%%%%%%%%%%%%%%%%%%%

\end{document}